\documentclass{article}
\usepackage{spconf,amsmath,graphicx}
\usepackage{amsmath,amssymb}
\usepackage{bbm}
\usepackage[font=small,skip=4pt]{caption} % or font=footnotesize
\usepackage{enumitem}
\usepackage{makecell}
\usepackage{xcolor}
\usepackage[skip=3pt]{caption} % space above/below captions
\usepackage[compact]{titlesec} % or just \usepackage{titlesec}
\setlist{nosep, leftmargin=14pt}

\usepackage{mwe} % to get dummy images
\usepackage{eso-pic}

\title{SplitFed-CL: A Split Federated Co-Learning Framework for Medical Image Segmentation with Inaccurate Labels}

\name{Zahra Hafezi Kafshgari, Hadi Hadizadeh and Parvaneh Saeedi%
\sthanks{© 2026 IEEE. Personal use of this material is permitted. Permission from IEEE must be obtained for all other uses, in any current or future media, including reprinting/republishing this material for advertising or promotional purposes, creating new collective works, for resale or redistribution to servers or lists, or reuse of any copyrighted component of this work in other works.}}

\address{School of Engineering Science, Simon Fraser University, Burnaby, BC, Canada}

\begin{document}

\AddToShipoutPictureFG*{%
  \AtPageUpperLeft{%
    \raisebox{-0.35in}{%
      \makebox[\paperwidth][c]{%
        \small\itshape To appear in the Proceedings of the 2026 IEEE International Conference on Image Processing (ICIP 2026)%
      }%
    }%
  }%
}

\maketitle

\begin{abstract}

Split Federated Learning (SplitFed) combines federated and split learning to preserve privacy while reducing client-side computation. However, in medical image segmentation, heterogeneous label quality across clients can significantly degrade performance.
We propose SplitFed-CL, a co-learning framework where a global teacher guides local students to detect and refine unreliable annotations. Reliable labels supervise training directly, while unreliable labels are corrected via weighted student--teacher refinement. SplitFed-CL further incorporates consistency regularization for robustness to input perturbations and a trainable weighting module to balance loss terms adaptively. We also introduce a novel difficulty guided strategy to simulate human like boundary centric annotation errors, where the degree of perturbation is governed by shape complexity and the associated annotation difficulty.
Experiments on two multiclass segmentation datasets with controlled synthetic noise, together with a binary segmentation dataset containing real-world annotation errors, demonstrate that SplitFed-CL consistently outperforms seven state-of-the-art baselines, yielding improved segmentation quality and robustness.

\end{abstract}


\section{Introduction}

Split Federated Learning (SplitFed) is a decentralized paradigm that integrates Split Learning (SL)~\cite{Split_Learning} and Federated Learning (FL)~\cite{fedavg} to achieve privacy-preserving, resource-efficient model training. In FL, clients retain data locally but must train full local models and transmit parameters for aggregation, which can require substantial on-device compute~\cite{guan2024federated}. SL reduces this load by splitting the network between clients and server, exchanging intermediate activations instead of raw data \cite{Split_Learning}. SplitFed combines these strengths: data never leave the clients while the server executes the heavy layers, enabling participation by resource-constrained devices~\cite{splitfed}.

In medical image segmentation, distributed datasets often suffer from inconsistent label quality due to varying annotator expertise \cite{survey_noisy_label}. These inaccuracies degrade the global model, especially in decentralized settings \cite{guan2024federated}.

In this paper, we propose SplitFed-CL, a co-learning framework for multiclass medical image segmentation that improves SplitFed training under heterogeneous and noisy annotations. Each client trains a student model guided by a global teacher that estimates label reliability. Reliable samples are used to directly update the model, while unreliable samples are refined through student--teacher predictions and incorporated with adaptive weighting. To systematically evaluate robustness under realistic annotation noise, we further introduce a difficulty-guided deformation strategy for simulating human-like segmentation errors. Moreover, we employ a consistency loss to enforce prediction invariance under input perturbations, together with a trainable loss-weighting module that automatically balances the reliable, unreliable, and consistency objectives, thereby eliminating manual tuning. Our main contributions are:

\begin{enumerate}
    \item A difficulty-guided framework for simulating human-like annotation errors.
    \item A global confidence-based mechanism for identifying reliable and unreliable annotations across clients.
    \item A student--teacher strategy for refining unreliable labels.
    \item A trainable loss-weighting module that automatically optimizes component contributions.
\end{enumerate}

\section{Related Work}

%Annotation inaccuracies are increasingly recognized as a fundamental robustness challenge in FL~\cite{uddin2025systematic}. Nevertheless, standardized evaluation protocols for FL under annotation noise remain limited~\cite{FedNoisy,fedBench}. While 

Several FL methods proposed to mitigate the effect of annotation noise for \emph{classification} tasks. For example, FedLN~\cite{FedLN} improves robustness through interpolation-based regularization and energy-driven scoring, while FedNoiL~\cite{FedNoil} identifies reliable clients and corrects labels using global model predictions. FedCorr~\cite{fedcorr} further incorporates intrinsic dimensionality estimation and Gaussian Mixture Models to detect and revise noisy labels, and FNBench~\cite{fedBench} offers a benchmarking framework for evaluating FL algorithms under label noise. FedNCL~\cite{FedNCL} introduces noise-robust aggregation by weighting clients based on sample reliability, while ARFL~\cite{ARFL} dynamically adjusts client influence via data quality. 

%However, extending such strategies to medical image segmentation is non-trivial, since segmentation relies on dense pixel-wise supervision and is particularly sensitive to boundary ambiguity and spatially correlated errors.
However, label correction in classification does not directly extend to segmentation, especially in medical imaging, where fuzzy boundaries hinder annotation accuracy.
Recent works tackle limited supervision and noisy data in FL segmentation. FedMix~\cite{fedmix} combines labeled and unlabeled data using consistency regularization and mix-up. FedA$^3$I~\cite{fedA3I} explicitly integrates annotation quality into aggregation. FedDM~\cite{fedDM} mitigates weak supervision and gradient conflicts through calibration and de-conflicting strategies. QA-SplitFed~\cite{Quality_adaptiveSplitFed} addresses inaccurate annotations in split federated learning by adaptively weighting client contributions based on estimated label quality. Lastly, DHLC and CELC~\cite{DHLC_CELC} refine lesion labels via student–teacher models, with CELC updating the global model using teacher parameters. 

In addition, most prior works simulate label noise using simplistic perturbations such as random label shuffling or basic contour dilation/erosion~\cite{ARFL, Quality_adaptiveSplitFed, DHLC_CELC}; in contrast, we propose a novel human-like annotation error method that generates realistic boundary deformations guided by a contour-based difficulty map.

\section{Simulating Annotation Error using a Difficulty Map}
\label{sec:difficulty_map_simulation}

To simulate human-like annotation errors in segmentation masks, we generate deformed version of accurate labels guided by a \emph{difficulty map} that emphasizes boundary regions where annotators are more likely to disagree. Intuitively, pixels with a higher probability of misannotation receive higher difficulty scores.

Given an input image $x$ and its corresponding mask $y$, we compute the signed distance function (SDF) $\phi$ from $y$ (negative inside the object and positive outside), and restrict all computations to a narrow boundary band as $B(u)=\mathbbm{1}\big(|\phi(u)| \le w\big)$ where $u$ denotes a pixel location and $w$ is the band width. We set $w = 0.2\sqrt{\frac{A_{\mathrm{obj}}}{\pi}}$, where $A_{\mathrm{obj}}$ is the object area in pixels and $\sqrt{\frac{A_{\mathrm{obj}}}{\pi}}$ is the equivalent radius, enabling scale-adaptive boundary modeling.

\paragraph{Boundary difficulty cues:}
Within $B(u)$, we estimate three cues that capture boundary uncertainty:
\begin{itemize}
      \item \textbf{Edge cue ($D_{\text{edge}}$):} Weak edges are harder to delineate consistently. We compute the normalized gradient magnitude as $g(u)=\left\lVert \nabla x(u)\right\rVert \in (0,1)$,
      where $\nabla$ denotes the spatial derivative operator~\cite{PratondoChuiOng2016}, and define the edge-based difficulty as
      \begin{equation}
      D_{\text{edge}}(u)=\big(1-g(u)\big)\, B(u)
      \end{equation}
   Therefore, strong edges yield low difficulty scores, whereas weak edges are assigned higher scores.

    \item \textbf{Blur cue ($D_{\text{blur}}$):} Reduced sharpness along object contours increases annotation uncertainty. We quantify local sharpness using the normalized magnitude of the Laplacian operator,
    $\textit{Laplacian}(u)=\left| \nabla^2 x(u)\right|  \in (0,1)$,
    where $\nabla^2$ denotes the second-order spatial derivative~\cite{Crete2007}, and define
    \begin{equation}
        D_{\text{blur}}(u)=\big(1-\textit{Laplacian}(u)\big)\,B(u)
    \end{equation}
    where higher difficulty values are assigned to more blurred boundary regions.

    \item \textbf{Curvature cue ($D_{\text{curv}}$):} Boundary points with high-curvatures are more prone to inconsistent delineation. Using the SDF, normalized curvature is approximated as $\kappa(u)\approx \nabla \cdot \left(\frac{\nabla \phi(u)}{\|\nabla \phi(u)\|}\right)  \in (0,1)$ following~\cite{ma2020learning}, and we define
    \begin{equation}
    D_{\text{curv}}(u)=|\kappa(u)|\,B(u)
    \end{equation}
    Accordingly, sharply changing contours are assigned higher difficulty than near-flat boundary segments.
\end{itemize}

\paragraph{Combined difficulty map ($D(u)$):}
We combine the cues into a normalized difficulty map $D(u)\in[0,1]$:
\begin{equation}
D(u)=\frac{D_{\text{edge}}(u)+D_{\text{blur}}(u)+D_{\text{curve}}(u) }{3} 
\end{equation}

\paragraph{Deformation magnitude and direction:}
We map difficulty to a bounded deformation magnitude as:
\begin{equation}
A(u)=\Big(a_{\min}+(a_{\max}-a_{\min})D(u)^{\rho}\Big)\,B(u)
\end{equation}
Where $\rho$ emphasizes high-difficulty boundary regions. Parameters $a_{\min}$, $a_{\max}$ and $\sigma$ controls the deformation rate. In this work, we set $\rho=2$ to emphasize high-difficulty boundaries, $a_{\min}=0$ and  $a_{\max}=w$.

For the deformation direction, we compute the outward unit normal 
$\mathbf{n}(u)=\nabla\phi(u)/\|\nabla\phi(u)\|$ and compare the edge evidence across the contour, where 
$g_{\mathrm{out}}(u)=g\!\big(u+\delta\,\mathbf{n}(u)\big)$ measures the edge strength outside the boundary and 
$g_{\mathrm{in}}(u)=g\!\big(u-\delta\,\mathbf{n}(u)\big)$ measures the edge strength inside the boundary. 
The signed displacement field is then defined as:
\begin{equation}
b(u)=\operatorname{clip}\!\left(\frac{g_{\mathrm{in}}(u)-g_{\mathrm{out}}(u)}{g_{\mathrm{in}}(u)+g_{\mathrm{out}}(u)+\varepsilon},-1,1\right),
\end{equation}
where $\varepsilon$ is a small constant added for numerical stability.
%indicating the more plausible boundary side: 
Following $b(u)$, if $g_{\mathrm{in}}(u)>g_{\mathrm{out}}(u)$, stronger evidence lies inside and the boundary is encouraged to shrink, whereas if $g_{\mathrm{in}}(u)<g_{\mathrm{out}}(u)$, stronger evidence lies outside and the boundary is encouraged to expand.

%if $g_{\mathrm{in}}(u)>g_{\mathrm{out}}(u)$, the contour is encouraged to shrink; otherwise, it expands.

\paragraph{Deformed mask:}
We deform the SDF and threshold it to obtain the noisy annotation as:
\begin{equation}
\small
\phi'(u)=\phi(u)+\big(G_{\sigma}\ast (A\,b)\big)(u), 
\quad
y_{\text{noisy}}=\mathbbm{1}\big(\phi'(u)\le 0\big), 
\end{equation} where $A(u)$ controls the deformation magnitude and $b(u)$ specifies its direction at pixel $u$. 
$G_{\sigma}$ denotes a Gaussian smoothing kernel with standard deviation $\sigma$, which suppresses unnaturally jagged boundaries. we set $\sigma=w$ in this study.
This pipeline generates realistic, boundary-centric annotation errors whose location, magnitude, and direction are guided by the image’s learned difficulty structure. For multi-class segmentation, each class is deformed using its own class-specific difficulty map, resulting in more realistic corrupted labels than uniformly deforming the entire mask. Representative examples of the resulting deformed labels are shown in Fig.~\ref{fig:results}, row~\textbf{(b)}.

\section{Proposed SplitFed Method}

\begin{figure}[!t]
  \centering
  \includegraphics[width=\columnwidth]{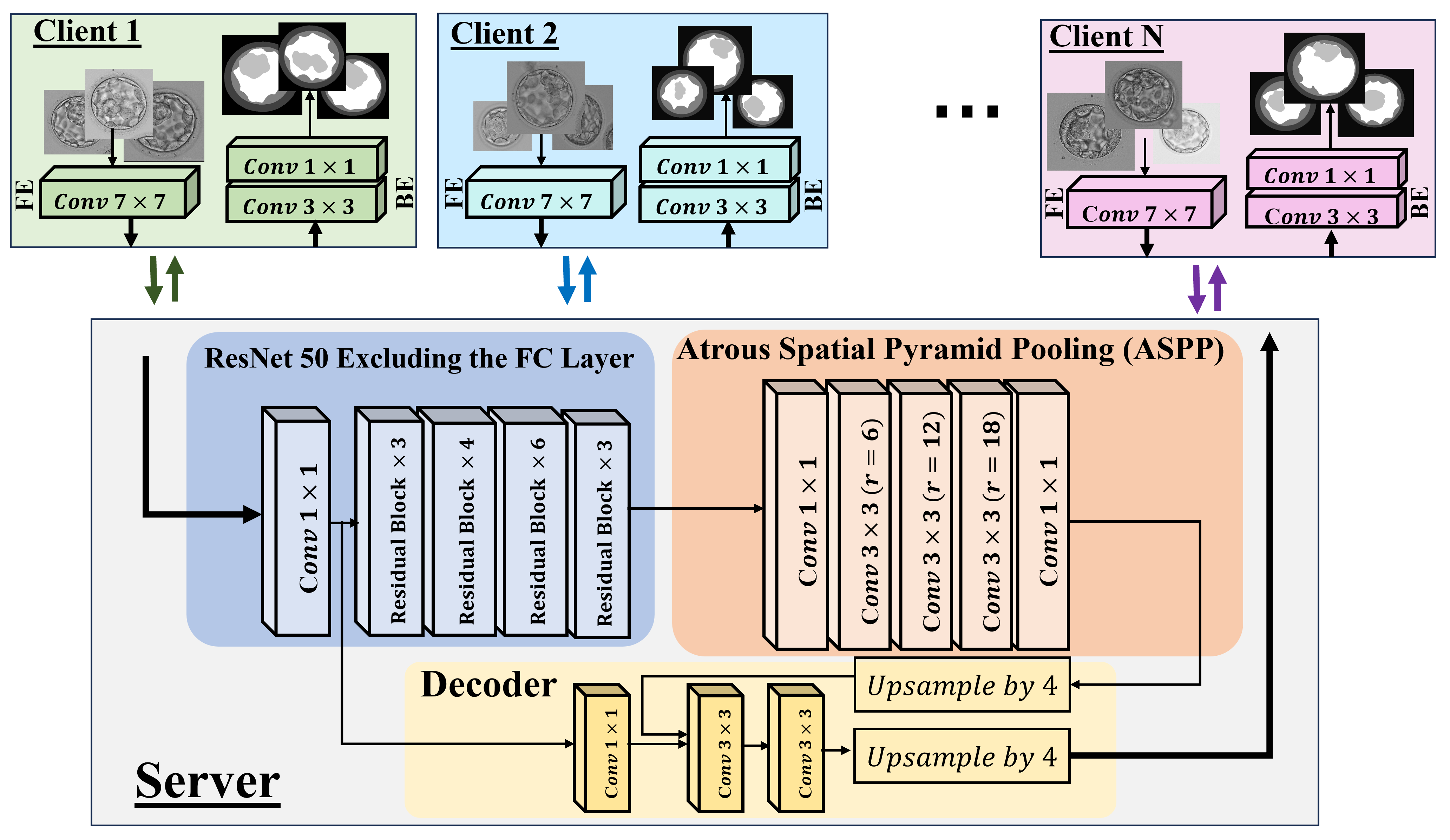}
  % or: \includegraphics[width=0.95\linewidth]{...}
  \caption{SplitFed DeepLabV3+ architecture: client-side FE/BE sub-models with a server-side sub-model for heavy computation. ResNet50~\cite{ResNet50} is used as the encoder backbone.}
  \label{fig:SplitFed_DeepLabV3Plus}
\end{figure}

We adopt DeepLabV3+~\cite{deeplabV3+} and partition it across client and server (Fig.~\ref{fig:SplitFed_DeepLabV3Plus}). On each client, a lightweight front-end ($\mathrm{FE}$) sub-model $f_{\theta_{\mathrm{FE}}}$ encodes the input and transmits activations to the server sub-model $f_{\theta_{\mathrm{S}}}$ for heavy computation; the returned feature maps are decoded by the client back-end ($\mathrm{BE}$) sub-model $f_{\theta_{\mathrm{BE}}}$ to produce the final prediction. This preserves data locality while offloading most compute to the server. At the start of each global round, the server distributes updated sub-models to clients; after local training, clients return updated weights for aggregation. We adopt a student–teacher setup: each client trains a student $\mathcal{F}=\{f_{\theta_{\mathrm{FE}}},f_{\theta_{\mathrm{S}}},f_{\theta_{\mathrm{BE}}}\}$, while the globally averaged model serves as a teacher $\bar{\mathcal{F}}=\{\bar f_{\theta_{\mathrm{FE}}},\bar f_{\theta_{\mathrm{S}}},\bar f_{\theta_{\mathrm{BE}}}\}$ that guides label reliability and provides a stable target for consistency.

For a batch $X=\{x_k\}_{k=1}^{K}$ with labels $Y=\{y_k\}_{k=1}^{K}$, the student and teacher produce predictions $f_{\theta_{\mathrm{BE}}}(X)$ and $\bar f_{\theta_{\mathrm{BE}}}(X)$ and per-sample region losses $\mathcal{L}_R=[\ell_1,\dots,\ell_K]$, $\bar{\mathcal{L}}_R=[\bar\ell_1,\dots,\bar\ell_K]$. Using a global threshold $\tau$, define the index sets
\begin{equation}
\small
\begin{aligned}
X_{\mathrm{re}},Y_{\mathrm{re}} &= \{\,x_k\in X , y_k\in Y \mid \ell_k\le \tau\ \text{and}\ \bar\ell_k\le \tau \,\},\\
X_{\mathrm{un}},Y_{\mathrm{un}} &= \{\,x_k\in X , y_k\in Y \mid \ell_k> \tau\ \text{and}\ \bar\ell_k> \tau \,\}.
\end{aligned}
\label{eq:reliable_unreliable_split}
\end{equation}

As detailed in Sec.~\ref{sec:tau}, $\tau$ starts high (most samples treated as reliable) and is updated each round using cross-client statistics and performance.

Unreliable labels $Y_{\mathrm{un}}$ are progressively refined to $\tilde Y_{\mathrm{un}}$ using student–teacher predictions (Sec.~\ref{sec:label_correction}), so they contribute proportionally while limiting noise. Robustness is further improved via a co-training–style consistency term that encourages the student on perturbed inputs $X'$ to match the teacher on original inputs $X$. The student is trained with:
\vspace{-4pt}
\begin{equation}
\small
\begin{aligned}
\mathcal{L}_{\mathrm{total}}
&= \mathcal{L}_{R}\!\big(f_{\theta_{\mathrm{BE}}}(X_{\mathrm{re}}), Y_{\mathrm{re}}\big)
+ \alpha\,\mathcal{L}_{R}\!\big(f_{\theta_{\mathrm{BE}}}(X_{\mathrm{un}}), \tilde Y_{\mathrm{un}}\big) \\
&\quad + \beta\,\mathcal{L}_{C}\!\big(f_{\theta_{\mathrm{BE}}}(X'), \tilde f_{\theta_{\mathrm{BE}}}(X)\big),
\end{aligned}
\label{eq:L_total}
\vspace{-6pt}
\end{equation}
where $\mathcal{L}_R$ is the region loss (applied to reliable and refined–unreliable labels) and $\mathcal{L}_C$ is a consistency term~\cite{co_training,consistency}. The coefficients $\alpha$ and $\beta$ automatically weight these components (see Sec.~\ref{sec:parameters}). 
After each local update on the student, the teacher sub-models are updated via an exponential moving average (EMA) of the student parameters, yielding a stable and progressively refined guidance signal~\cite{co_training}.

\subsection{SplitFed Averaging Method}
\label{sec:aggregation}

After local training, each client transmits its student parameters together with summary statistics (total samples, reliable-sample count, and reliable-subset loss) to the server. This enables reliability-aware aggregation that accounts for both data quantity and quality.

Let $d_{\mathrm{re}}^{i}$ be the number of reliable samples for client $i=1,..,N$ and $\mathcal{L}_{\mathrm{re}}^{i}$ its mean reliable loss. Define the data ratio $\mathbf{d}=[d_{\mathrm{re}}^{i}]_{i=1}^{N}\big/\sum_{j=1}^{N} d_{\mathrm{re}}^{j}$ and the performance ratio $\mathbf{q}=\operatorname{softmax}(-\gamma\,\mathbf{L})$ with $\mathbf{L}=[\mathcal{L}_{\mathrm{re}}^{i}]_{i=1}^{N}$, where $\gamma$ is initialized small and increased over global epochs to emphasize lower-loss clients. The contribution ratios are finally computed as $\mathbf{r}=\frac{\mathbf{q}\odot \mathbf{d}}{\mathbf{q}^{\top}\mathbf{d}} = [\,r_1,\dots,r_N\,]^{\top}.$\
Using $\mathbf{r}$, global sub-model parameters are updated via a reliability-weighted average:
\vspace{-2pt}
\begin{equation}
\small
\{\overline{\theta}_{\mathrm{FE}},\,\overline{\theta}_{\mathrm{S}},\,\overline{\theta}_{\mathrm{BE}}\}
\;\gets\;
\left\{
\sum_{i=1}^{N} r_i\,\theta_{\mathrm{FE}}^{i},\;
\sum_{i=1}^{N} r_i\,\theta_{\mathrm{S}}^{i},\;
\sum_{i=1}^{N} r_i\,\theta_{\mathrm{BE}}^{i}
\right\}
\label{eq:aggregation}
\end{equation}
\vspace{-2pt}

The updated teacher and student parameters are then broadcast for the next round. By weighting clients based on the amount and quality of \emph{reliable} data, rather than total volume, this aggregation reduces the influence of inaccurate annotations and improves global performance.

\subsection{Updating Global Threshold}
\label{sec:tau}

\noindent
Motivated by the intuition that reliable samples can be identified through their training-loss behavior, we update the global threshold as a reliability-weighted statistic, $\tau = \mathbf{b}^{\top}\mathbf{q}$ where $\mathbf{b}=\big[\mu_1+\lambda\sigma_1,\dots,\mu_N+\lambda\sigma_N\big]$ and $\mathbf{q}$ is the performance-ratio vector (Sec.~\ref{sec:aggregation}) and $\mu_i,\sigma_i$ denote the mean and standard deviation of per-sample training losses for client $i$.
The scalar $\lambda>0$ controls the upper bound of this confidence statistic and is decayed over global epochs, making $\tau$ increasingly selective.

Leveraging reliable-sample identification, $\tau$ uses $\mathbf{q}$ as reliability weights to apply the client-specific bounds in $\mathbf{b}$ when filtering all samples.
Initially, $\tau$ is set to a high value (e.g., 10) so local models leverage most data; as training proceeds and $\lambda$ shrinks, $\tau$ tightens to focus learning on more reliable samples, improving robustness.

\subsection{Label Correction}
\label{sec:label_correction}

After identifying reliable and unreliable labels in $Y$, unreliable labels, $Y_{un}$, are locally modified to $\bar{Y}_{un}$ using predictions from both the student and teacher models. 
A difference mask $R$ is generated for each unreliable label to highlight areas of uncertainty between the predictions of the student model, teacher model, and the ground truth label, defined by $\small
R = P_{\mathrm{un}} \triangle \bar{P}_{\mathrm{un}} \triangle Y_{\mathrm{un}}$. Here, $\triangle$ is the symmetric difference, $P= \operatorname*{arg\,max}_{c \in \{1,\dots,C\}} f_{\theta_{\mathrm{BE}}}(X_{un})$ and $\overline{P}= \operatorname*{arg\,max}_{c \in \{1,\dots,C\}} \bar{f}_{\theta_{\mathrm{BE}}}(X_{un})$ are the hard-labeled masks with $C$ segmentation classes. 
Subsequently, the modified $Y_{un}$ is computed as follows:

\begin{equation}
\small
\tilde{Y}_{un}(D) =
\begin{cases} 
P_{un}(R) & \text{if } f_{\theta_{\mathrm{BE}}}(X_{un}) > T \\ 
\bar{P}_{un}(R) & \text{if } \bar{f}_{\theta_{\mathrm{BE}}}(X_{un}) > T \\
Y_{un} & \text{otherwise}
\end{cases}.
\label{eq:labeling} 
\end{equation}

The modified labels $\tilde{Y}_{un}$ are then utilized as the ground truth to compute the region loss, as described in this section.

% \subsection{Adaptive Loss Weighting (Trainable Coefficients)}
\subsection{Learning Loss Coefficients}
\label{sec:parameters}

As defined in Eq.~\eqref{eq:L_total}, the coefficients $\alpha$ and $\beta$ weight the unreliable-subset region loss and the consistency loss, respectively. Manual tuning is costly and brittle. Inspired by Kendall–Gal~\cite{kendall2018multi}, we replace fixed coefficients with a trainable weighting module and optimize them during training. Concretely, we use:
\begingroup
\small
\vspace{-4pt}
\begin{equation}
\begin{aligned}
\mathcal{L}_{\mathrm{total}}
&= w_{\mathrm{re}}\,\widehat{\mathcal{L}}_{R}\!\big(f_{\theta_{\mathrm{BE}}}(X_{\mathrm{re}}), Y_{\mathrm{re}}\big)
+ w_{\mathrm{un}}\,\widehat{\mathcal{L}}_{R}\!\big(f_{\theta_{\mathrm{BE}}}(X_{\mathrm{un}}), \tilde{Y}_{\mathrm{un}}\big) \\
&\quad + w_{\mathrm{cons}}\,\widehat{\mathcal{L}}_{C}\!\big(f_{\theta_{\mathrm{BE}}}(X'), \tilde f_{\theta_{\mathrm{BE}}}(X)\big)
+ \mathcal{R}(u),
\end{aligned}
\label{eq:L_total_new}
\end{equation}
\endgroup
%\vspace{-4pt}
with $w_{\mathrm{re}}=1$ and $w_{un}(t) = \sigma\!(u_{\mathrm{un}})\,s(t)\in(0,1)$ and $w_{cons}(t) = \sigma\!(u_{\mathrm{cons}})\,s(t)\in(0,1)$. Here, $\sigma(\cdot)$ is the  logistic sigmoid and $s(t)$ is a linear warm-up schedule that keeps $w_{\mathrm{un}}$ and $w_{\mathrm{cons}}$ small in early epochs. The logits $u_{\mathrm{un}}$ and $u_{\mathrm{cons}}$ are trainable; the terms $\widehat{\mathcal{L}}$ are EMA-normalized to maintain comparable, stable scales. We initialize $u_{\mathrm{un}},u_{\mathrm{cons}}\ll 0$ (e.g., $-10$), so both weights start near zero $w_{\text{cons}},w_{\text{noisy}}\approx 0$ and are allowed to grow toward $1$ as training progresses. To mitigate overconfident saturation, we include a small $\ell_2$ regularizer,
$\mathcal{R}(u)=\eta\left(u_{\mathrm{un}}^{2}+u_{\mathrm{cons}}^{2}\right)$.
At each global round, local logits $u_{\mathrm{un}}$ and $u_{\mathrm{cons}}$ are aggregated using the reliability-weighted averaging rule described in Sec.~\ref{sec:tau}. In all experiments, we set $\eta=5\times 10^{-4}$.

\section{Experiments}
\subsection{Datasets}
\label{sec:Datasets}

To evaluate the effectiveness of the proposed SplitFed-CL method, we utilize two multi-class and one binary medical image segmentation dataset.
The \textit{Human Embryo} dataset comprises 594 microscopic images of human blastocysts, with labels segmenting four structures: Zona Pellucida (ZP), Trophectoderm (TE), Inner Cell Mass (ICM), and Blastocoel (BL)~\cite{saeedi2017automatic}. In the SplitFed setup, the data are distributed heterogeneously across four clients (100, 150, 200, and 50 images, respectively), while 94 precisely annotated images are reserved for testing. The second dataset, Pubic Symphysis and Fetal Head Segmentation (\textit{PSFHS})~\cite{PSFHS}, introduced in the MICCAI 2023 challenge, contains 1,358 annotated intrapartum transperineal ultrasound images for segmenting the pubic symphysis (PS) and fetal head (FH). Although newer samples have been released, we use only the initial subset due to its higher annotation reliability. The PSFHS data are allocated across four clients (100, 200, 450, and 250 images, respectively) with 358 images retained for evaluation. 
To model heterogeneous annotation quality, we corrupt a subset of each client’s labels using the deformation strategy in Sec.~\ref{sec:difficulty_map_simulation}, with corruption ratios of 20\%, 50\%, 80\%, and 0\% for Clients~1--4, respectively.

To evaluate the proposed SplitFed-CL framework under a more realistic annotation noise condition, we included a third dataset, \textit{ISIC}, a dermoscopic skin-lesion segmentation benchmark from the \textit{ISIC} Archive~\cite{codella2018isic}. Specifically, we used ISIC MultiAnnot++ (IMA++), which contains 12,573 single-annotation masks and 2,394 multi-annotator masks with tool/skill metadata~\cite{abhishek2024segmentation, abhishek2025can}. 

Using the IMA++ metadata, we selected 600 multi-annotator samples annotated using a semi-automated flood-fill tool (T2) with expert-defined parameters and verified by a novice (S2), denoted as \textit{T2/S2}, and treated them as real-world unreliable annotations. We also selected 5,124 single-annotation masks generated via manual polygon tracing by an expert (T1) and verified by an expert (S1), denoted as \textit{T1/S1}, and treated them as accurate samples.
We distributed these samples across five clients containing 500, 800, 800, 1500, and 1000 images, respectively, while retaining 1,124 \textit{T1/S1} images for evaluation. The corruption ratios for clients~1--5 were 0\%, 50\%, 50\%, 80\%, and 60\%, respectively. The corrupted labels in client~5 corresponded to real-world noisy annotations (\textit{T2/S2}), whereas those in client~4 were synthetically generated using the proposed difficulty-guided mask deformation strategy (Sec.~\ref{sec:difficulty_map_simulation}).

%We then distribute 1,000 samples to each of five clients, with corruption ratios of 0\%, 20\%, 80\%, 50\%, and 60\% for clients~1--5, respectively. The corrupted labels in client~5 are real-world noisy annotations (\textit{T2/S2}), whereas corrupted labels in the other clients are generated using synthetic mask deformations (Sec.~\ref{sec:difficulty_map_simulation}). 

%To simulate varying data quality, a subset of each client's annotations is deliberately corrupted by randomly eroding or dilating the accurate labels' components using circular or square structuring elements. The radius of these elements varies randomly between 5 and 15. For example, client $i$'s dataset consists of $a$\% inaccurate and $(100-a)$\% accurate labels to reflect the impact of heterogeneity. Two scenarios with differing corruption rates are evaluated as:

%\begin{itemize}
%\small
%\item \textbf{Scenario 1}: 20\%, 50\%, 80\%, and 0\% corruption for Clients 1–4, respectively, and
%\item \textbf{Scenario 2}: 20\%, 80\%, 80\%, and 20\% corruption for Clients 1–4, respectively.
%\end{itemize}

\begin{table}[t]
  \caption{Performance comparison on \textbf{\textit{PSFHS}} dataset for \textbf{PS} \& \textbf{FH} segments. } 
  \centering
  \resizebox{\columnwidth}{!}{%
  \begin{tabular}{|l|c|c|c|c|c|}
    \hline
    \textbf{Method} & \textbf{Accuracy} & \textbf{Dice Loss} & \textbf{Mean IOU} & \textbf{PS IOU} & \textbf{FH IOU} \\
    \hline\hline
    All Accurate Labels (FedAVG~\cite{fedavg}) &  0.9635 & 0.0426 & 0.9605 & 0.8701 & 0.9196 \\ 
    All Accurate Labels (\textit{\textbf{SplitFed-CL}}) &  0.9810 & 0.0418 & 0.9630 & 0.8630 & 0.9290 \\ \hline 
    FedAVG~\cite{fedavg} &  0.9430  & 0.0585 & 0.9475 & 0.8191 & 0.8970 \\
    FedMix~\cite{fedmix} &  0.9412 & 0.0844 & 0.9430 & 0.6912 & 0.8984 \\
    FedNCL-V2~\cite{FedNCL} & 0.9326   & 0.0793 & 0.9456 & 0.7065 & 0.9055 \\
    ARFL~\cite{ARFL} &  0.9312  & 0.0856 & 0.9420 & 0.6913 & 0.8920 \\
    QA-SplitFed~\cite{Quality_adaptiveSplitFed} & 0.9578   & 0.0574 & 0.9493 & 0.8235 & 0.8982 \\
    CELC~\cite{DHLC_CELC} &  0.9127  & 0.1620 & 0.8701 & 0.5488 & 0.7547 \\
    DHLC~\cite{DHLC_CELC} &  0.9111  & 0.1650 & 0.8625 & 0.5570 & 0.7339 \\ \hline 
    \textit{\textbf{SplitFed-CL}} (No Label Correction) &  0.9715  & 0.0425 & 0.9537 & 0.8561 & 0.9330 \\
    \textit{\textbf{SplitFed-CL}} (No Consistency Loss) &  0.9725  & 0.0399 & 0.9556 & 0.8660 & 0.9330 \\
     \textit{\textbf{SplitFed-CL}} (full) &  \textbf{0.9788}  & \textbf{ 0.0363 } & \textbf{ 0.9592 } & \textbf{ 0.8805 } & \textbf{  0.9424 } \\
    \hline
  \end{tabular}}
  \label{tab:Fetal_Scenario_1}
\end{table}

% \begin{table}[t]
  %\caption{Performance comparison on \textit{PSFHS} (Scenario 2) for Pubic Symphysis and Fetal Head segments (PS \& FH). Top row is the clean-label baseline.} 
 % \centering
%  \resizebox{0.95\columnwidth}{!}{%
 % \begin{tabular}{|l|c|c|c|c|}
  %  \hline
  %  \textbf{Method} & \textbf{Dice Loss}~\ensuremath{\downarrow} & \textbf{Mean IOU}~\ensuremath{\uparrow} & \textbf{PS IOU}~\ensuremath{\uparrow} & \textbf{FH IOU}~\ensuremath{\uparrow} \\
  %  \hline\hline
  %  All Accurate Labels &  0.0426 & 0.9605 & 0.8701 & 0.9196 \\ \hline 
  %  FedAVG~\cite{fedavg}   &    &     &   &     \\ 
  %  FedMix~\cite{fedmix}  &     &       &        &         \\ 
  %  FedNCL-V2~\cite{FedNCL} &     &       &        &         \\ 
  %  ARFL~\cite{ARFL} &     &       &        &         \\ 
  %  QA-SplitFed~\cite{Quality_adaptiveSplitFed} &     &    &    &    \\ 
 %   CELC~\cite{DHLC_CELC} &     &       &        &         \\ 
  %  DHLC~\cite{DHLC_CELC} &     &       &        &         \\ 
   %    \textit{\textbf{SplitFed-CL} V4} &  \textbf{ 0.0381 } & \textbf{ 0.9662} & \textbf{ 0.8738  } & \textbf{ 0.9348} \\
   %    \textit{\textbf{SplitFed-CL} V5} &  \textbf{ 0.0344 } & \textbf{ 0.9597 } & \textbf{  0.8920 } & \textbf{ 0.9412} \\
  %  \hline
 % \end{tabular}}
  %\label{tab:Fetal_Scenario_1}
% \end{table}

\subsection{Results and Discussion}

We evaluated our SplitFed method against seven state-of-the-art baselines adapted to the SplitFed setting: FedAvg~\cite{fedavg} (server aggregation $\overline{\mathbf W}=\big(\sum_{i=1}^{N} m_i\big)^{-1}\sum_{i=1}^{N} m_i\,\mathbf W_i$, with $m_i$ the client sample counts); FedMix~\cite{fedmix} ($\beta=1.5$, $\lambda=10$); FedNCL-V2~\cite{FedNCL} ($\alpha=\beta=2$); ARFL~\cite{ARFL} (with $M_P=\lambda$ as in the original); Quality-Adaptive SplitFed (QA-SplitFed)~\cite{Quality_adaptiveSplitFed} (aggregation tailored for noisy clients); and two label-correction methods, DHLC and CELC~\cite{DHLC_CELC}.

To our knowledge, QA-SplitFed~\cite{Quality_adaptiveSplitFed} is the only SplitFed framework explicitly targeting noisy labels. Accordingly, for a fair comparison we ported the remaining FL baselines to the SplitFed protocol (same client–server partition, communication schedule, and comparable hyperparameter budgets).

Input images are resized to $352 \times 352$, and augmentation is performed using horizontal and vertical flipping, and rotation of up to $\pm35^{\circ}$. The system is trained with the Adam optimizer with the learning rate set to $10^{-4}$.
All models are trained for 5 local epochs per round and validated over 100 global epochs. Threshold $T$ in Equation~\ref{eq:labeling} is set to 0.9. 
%For model comparison, we utilized established evaluation metrics based on True Positive (TP), True Negative (TN), False Positive (FP), and False Negative (FN) pixels as $\text{Dice loss}= 1-\frac{2TP}{2TP+FP+FN}$, and $\text{IOU}= \frac{TP}{TP+FP+FN}$.
For evaluation, we report pixel-wise Accuracy ({\small$\frac{TP+TN}{TP+TN+FP+FN}$}), Dice loss ({\small$1-\frac{2TP}{2TP+FP+FN}$}), and IoU ({\small$\frac{TP}{TP+FP+FN}$}) computed from (\small$TP,TN,FP,FN$) for \textit{Human Embryo} and \textit{PSFHS}. 
For \textit{ISIC}, we additionally report Precision ({\small$\frac{TP}{TP+FP}$}) and Recall ({\small$\frac{TP}{TP+FN}$}).

We define $\alpha=w_{\mathrm{un}}(t)$ and $\beta=w_{\mathrm{cons}}(t)$ with trainable logits $u_{\mathrm{un}},u_{\mathrm{cons}}$. The logits are optimized with Adam (lr $=10^{-3}$) and
initialized as $u_{\mathrm{un}}=u_{\mathrm{cons}}=-10$, so
$w_{\mathrm{un}}(0),w_{\mathrm{cons}}(0)\approx 0$. 
The temperature parameter $\gamma$ in Sec.~\ref{sec:aggregation} is linearly warmed up from $1$ to $5$ over the first $50$ global epochs, while
$\lambda$ in Sec.~\ref{sec:tau} decays linearly from $3$ to $0$ over the same period and then kept constant thereafter.

Tables~\ref{tab:Fetal_Scenario_1}--\ref{tab:ISIC} report quantitative results on \textit{PSFHS}, \textit{Human Embryo}, and \textit{ISIC}. In each table, the first two rows represent the clean-label reference setting (all clients have accurate annotations), using FedAvg and \textit{\textbf{SplitFed-CL}} for aggregation, respectively. For \textit{ISIC}, this baseline is obtained by replacing \textit{T2/S2} annotations with the corresponding \textit{T1/S1} masks. Best results are shown in bold.
Across all datasets, SplitFed-CL consistently surpasses prior SOTA methods and achieves performance closest to the clean-label reference. Ablation results further show that removing either label correction or consistency loss degrades performance, confirming the importance of both components.

\begin{figure}[!t]
  \centering
  \includegraphics[width=\columnwidth]{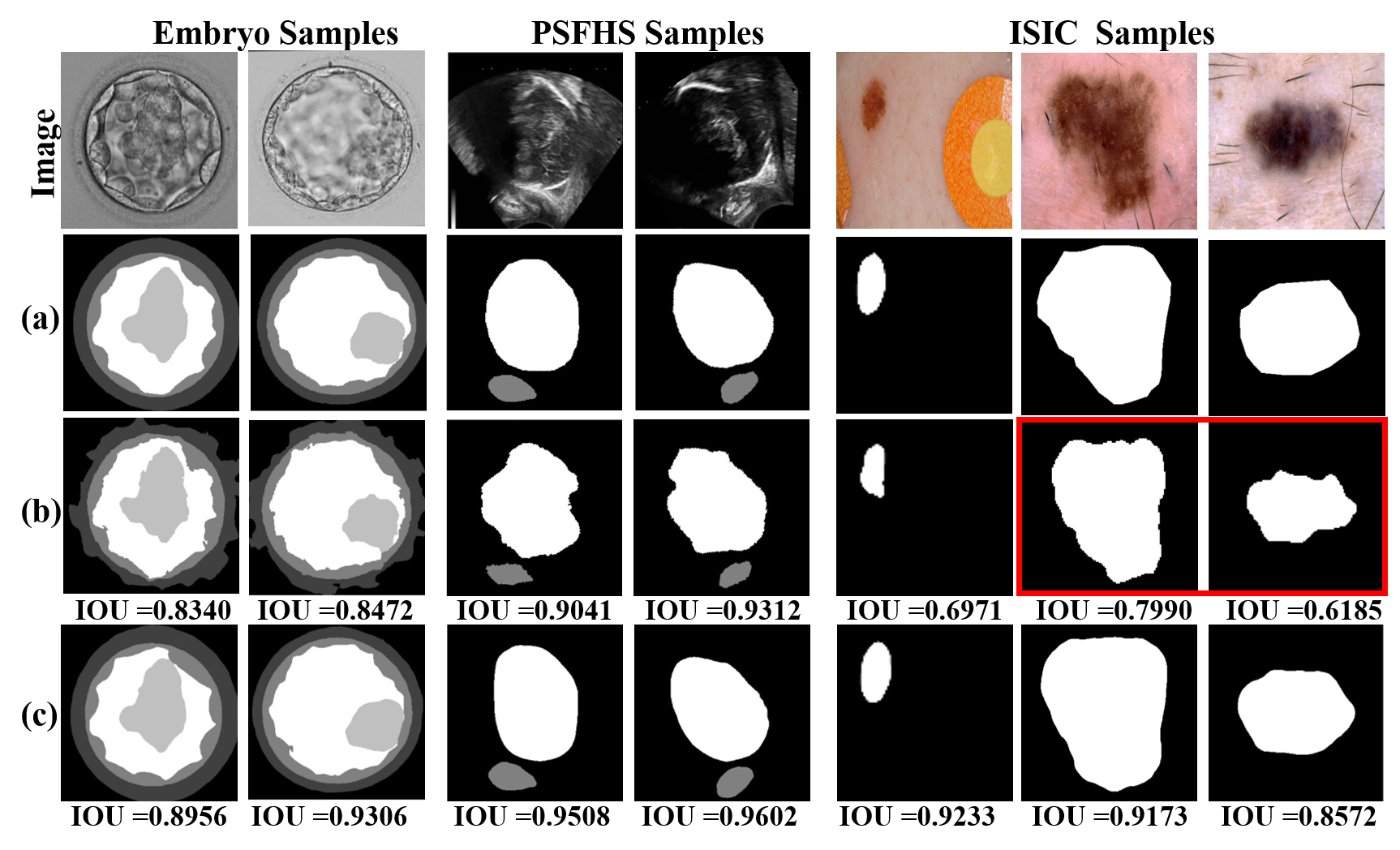}
  % or: \includegraphics[width=0.95\linewidth]{...}
  \caption{Qualitative label-correction results on representative samples from three datasets under the proposed SplitFed-CL framework, showing \textbf{(a)} accurate, \textbf{(b)} inaccurate, and \textbf{(c)} corrected labels. Labels in the \textcolor{red}{red} box correspond to real-world \textit{ISIC} T2/S2 annotations.}

 % : input image samples; \textbf{(b)}: accurate labels $y$; \textbf{(c)}: inaccurate labels $y_{un}$; \textbf{(d)}: the modified labels $\tilde{y}_{un}$.}
  \label{fig:results}
\end{figure}

To demonstrate the impact of label correction, Fig.~\ref{fig:results} demonstrates some examples of accurate, corrupted, and modified labels from the three datasets. The mean IoU is calculated for corrupted and corrected labels against the accurate labels. For the last samples highlighted by the red box, corrupted labels are real \textit{ISIC} \textit{T2/S2} annotations; the rest of noisy labels are synthetically generated as described in Sec.~\ref{sec:difficulty_map_simulation}. As shown in Fig.~\ref{fig:results}, SplitFed-CL refines corrupted labels, improving local training and ultimately enhancing the global model, consistent with Tables~\ref{tab:Fetal_Scenario_1}--\ref{tab:ISIC}.

\begin{table}[t]
  \caption{Performance comparison on \textbf{\textit{Human Embryo}} dataset for \textbf{ZP}, \textbf{TE}, \textbf{ICM} and \textbf{BL} segments.}
  \centering
  \resizebox{\columnwidth}{!}{%
  \begin{tabular}{|l|c|c|c|c|c|c|c|}
    \hline
    \textbf{Method} &
    \makecell{\textbf{Accuracy}} &
    \makecell{\textbf{Dice}\\\textbf{Loss} } &
    \makecell{\textbf{Mean}\\\textbf{IoU} } &
    \makecell{\textbf{ZP}\\\textbf{IoU} } &
    \makecell{\textbf{TE}\\\textbf{IoU} } &
    \makecell{\textbf{ICM}\\\textbf{IoU} } &
    \makecell{\textbf{BL}\\\textbf{IoU} } \\
    \hline\hline

    All Accurate Labels (FedAVG~\cite{fedavg})&  0.9368  &  0.0960 &  0.8661 &  0.7684 & 0.7339 & 0.8576 &  0.8605 \\ 
    All Accurate Labels (\textit{\textbf{SplitFed-CL}}) &  0.9448  &  0.0753 &  0.8960 &  0.8147 & 0.7816 & 0.8811 &  0.8898 \\ \hline 
    FedAVG~\cite{fedavg} &  0.9234 & 0.1074 & 0.8557 & 0.7883 & 0.6846 & 0.8372 & 0.8405 \\
    FedMix~\cite{fedmix} &   0.9289 & 0.1033 & 0.8608 & 0.7780 & 0.7047 & 0.8423 & 0.8514 \\
    FedNCL-V2~\cite{FedNCL} &   0.9420  & 0.0776 & 0.8907 & \textbf{0.8281} & 0.7828 & 0.8702 & 0.8814 \\
    ARFL~\cite{ARFL} &    0.9365 & 0.0830 & 0.8839 & 0.8212 & 0.7713 & 0.8573 & 0.8732 \\
    QA-SplitFed~\cite{Quality_adaptiveSplitFed} &  0.9312   & 0.0945 & 0.8695 & 0.8077 & 0.7318 & 0.8460 & 0.8632 \\
    CELC~\cite{DHLC_CELC} & 0.9178   &  0.1387 & 0.8209 & 0.7296 & 0.5846 & 0.8156 & 0.8315 \\
    DHLC~\cite{DHLC_CELC} &   0.9232  & 0.1349 & 0.8225 & 0.7205 & 0.6037 & 0.8185 & 0.8416 \\ \hline 
    \textit{\textbf{SplitFed-CL}} (No Label Correction) &  0.9326  & 0.0806 & 0.8825 & 0.8106 & 0.7533 & 0.8612 & 0.8655 \\
    \textit{\textbf{SplitFed-CL}} (No Consistency Loss) &  0.9434  & 0.0775 & 0.8933 & 0.8250 & 0.7651 & 0.8705 & 0.8799 \\
    \textit{\textbf{SplitFed-CL}} (full) &  \textbf{0.9451}   & \textbf{0.0725} & \textbf{0.8995} & 0.8265 & \textbf{0.7862} & \textbf{0.8780} & \textbf{0.8923} \\
    \hline
  \end{tabular}%
  }
  \label{tab:Embryo_Scenario_1}
\end{table}

\begin{table}[t]
  \caption{Performance comparison on \textit{\textbf{ISIC}} dataset. } 
  \centering
  \resizebox{\columnwidth}{!}{%
  \begin{tabular}{|l|c|c|c|c|c|}
    \hline
    \textbf{Method} & \textbf{Accuracy} & \textbf{Dice Loss}& \textbf{FG IOU} & \textbf{Precision}& \textbf{Recall} \\
    \hline\hline
    All Accurate Labels (FedAVG~\cite{fedavg})&  0.9860 &  0.1320 & 0.7641 & 0.9073 & 0.9156 \\ 
    All Accurate Labels (\textit{\textbf{SplitFed-CL}}) &  0.9850 &  0.1377 & 0.7580 & 0.8974 & 0.9149 \\ \hline 
    FedAVG~\cite{fedavg} & 0.9801  & 0.1616 &  0.7199 &  0.9044 &  0.8348 \\ 
    FedMix~\cite{fedmix} &  0.9811 & 0.1488 & 0.7326  &  0.8916 & 0.8467 \\ 
    FedNCL-V2~\cite{FedNCL} &  0.9822  & 0.1389 &  0.7401 & 0.8910  & 0.8532 \\ 
    ARFL~\cite{ARFL} & 0.9800  & 0.1411 &  0.7320 &  0.8901 & 0.8442 \\ 
    QA-SplitFed~\cite{Quality_adaptiveSplitFed} &  0.9827 &  0.1402 &  0.7432 &  0.9002 & 0.8732 \\ 
    CELC~\cite{DHLC_CELC} & 0.9555  & 0.1890 & 0.6912  & 0.8555  & 0.8321 \\ 
    DHLC~\cite{DHLC_CELC} & 0.9619  & 0.1853 &  0.7001 &  0.8584 & 0.8333 \\ \hline
    \textit{\textbf{SplitFed-CL}} (No Label Correction) &  0.9814  & 0.1544 &  0.7413 &  \textbf{0.9393} & 0.8432  \\
    \textit{\textbf{SplitFed-CL}} (No Consistency Loss) &  0.9826  & 0.1438 &  0.7544 &  0.9330 & 0.8665  \\
     \textit{\textbf{SplitFed-CL}} (full)&  \textbf{ 0.9830}  & \textbf{ 0.1320 } & \textbf{ 0.7641 } &  0.9074  & \textbf{  0.9157 } \\
    \hline
  \end{tabular}}
  \label{tab:ISIC}
\end{table}

\begin{figure}[!t]
  \centering
  \includegraphics[width=\columnwidth]{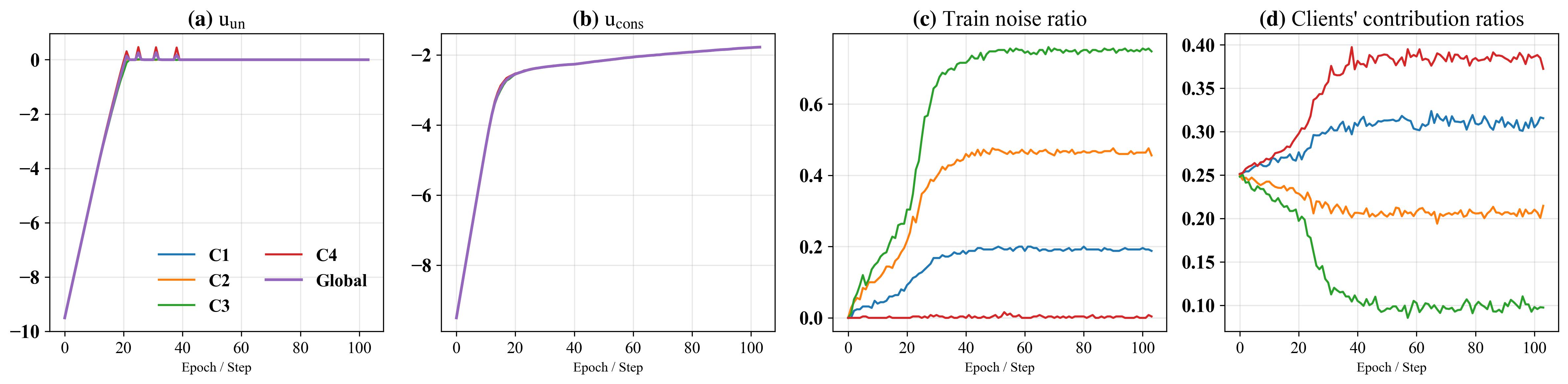}
  % or: \includegraphics[width=0.95\linewidth]{...}
  \caption{\textit{PSFHS}: Per-client training dynamics—\textbf{(a)} local $u_{un}$, \textbf{(b)} local $u_{cons}$, \textbf{(c)} train-noise ratio, \textbf{(d)} aggregation contribution; curves: C1 to C4}
  \label{fig:Evolution}
\end{figure}

%\noindent\textbf{Evolution of training signals.} 
Fig.~\ref{fig:Evolution} tracks client-level quantities on \textit{PSFHS} dataset. 
The learnable logits \(u_{\mathrm{un}}\) and \(u_{\mathrm{cons}}\) in Eq.~\eqref{eq:L_total_new} are optimized locally and aggregated per Sec.~\ref{sec:aggregation}. 
By the end of training, the global logits converge to \(u_{\mathrm{un}}\!\approx\!0\) and \(u_{\mathrm{cons}}\!\approx\!-1.77\), yielding
\(w_{\mathrm{un}}=\sigma(u_{\mathrm{un}})=0.5\) and 
\(w_{\mathrm{cons}}=\sigma(u_{\mathrm{cons}})\approx 0.1455\) for the unreliable-label and consistency losses, respectively.
The model’s estimated training-noise ratio (Fig.~\ref{fig:Evolution}c) closely matches the injected corruption: ground truth 20\%, 50\%, 80\%, 0\% for C1–C4 versus detected 18.8\%, 45.6\%, 74.8\%, 0.4\% using the threshold \(\tau\) from Sec.~\ref{sec:label_correction}. 
Finally, the reliability-aware aggregation adapts client contributions over rounds (Fig.~\ref{fig:Evolution}d): with \(\gamma\) initialized small, the ratios start similar and then shift toward lower-loss clients as \(\gamma\) increases (Sec.~\ref{sec:aggregation}).

%globally reached to 0 and 
%(1,1) Local w_un (noisy-loss weight) for clients C1–C4. (1,2) Local w_cons (consistency-loss weight) for C1–C4. (2,1) Estimated train noise ratio per client. (2,2) Clients’ contribution ratios used in joint aggregation. Each curve corresponds to one client; x-axis denotes epochs/rounds; y-axes are the metric values reported by the training pipeline. }

\section{Conclusion}

%We introduced SplitFed-CL, a Split Federated co-learning framework for medical image segmentation under noisy and inconsistent labels. The method integrates student–teacher learning with reliability-aware aggregation to identify, refine, and reweight unreliable annotations. A global confidence metric separates reliable samples, while label correction, consistency regularization, and adaptive loss weighting jointly enhance robustness and stability.
%Experiments on two medical segmentation datasets with controlled label noise show that SplitFed-CL consistently outperforms existing FL and SplitFed baselines, achieving performance close to clean-label training. These results highlight its effectiveness in improving model reliability and noise resilience in decentralized medical imaging.

We introduce SplitFed-CL, a Split Federated co-learning framework for medical image segmentation under noisy and inconsistent labels. The method integrates student–teacher learning with reliability-aware aggregation to identify, refine, and reweight unreliable annotations. A global confidence metric separates reliable samples, while label correction, consistency regularization, and adaptive loss weighting jointly enhance robustness and stability. Experiments on two medical segmentation datasets with controlled label noise show that SplitFed-CL consistently outperforms existing FL and SplitFed baselines, achieving performance close to clean-label training. These results highlight its effectiveness in improving model reliability and noise resilience in decentralized medical imaging.

\small
\bibliographystyle{IEEEbib}
\bibliography{refs}

@inproceedings{fedcorr,
  title={Fedcorr: Multi-stage federated learning for label noise correction},
  author={Xu, J. and Chen, Z. and Quek, T. QS. and Chong, K. F. E.},
  booktitle={Proceedings of the IEEE/CVF conference on computer vision and pattern recognition},
  pages={10184--10193},
  year={2022}
}

@article{PSFHS,
  title={PSFHS: intrapartum ultrasound image dataset for AI-based segmentation of pubic symphysis and fetal head},
  author={Chen, G. and Bai, J. and Ou, Zh. and Lu, Y. and Wang, H.},
  journal={Scientific Data},
  volume={11},
  number={1},
  pages={436},
  year={2024},
  publisher={Nature Publishing Group UK London}
}

@article{fedBench,
  title={Fnbench: Benchmarking robust federated learning against noisy labels},
  author={Jiang, X. and Li, J. and Wu, N. and Wu, Zh. and Li, X. and Sun, Sh. and Xu, G. and Wang, Y. and Li, Q. and Liu, M.},
  journal={arXiv preprint arXiv:2505.06684},
  year={2025}
}

@article{DHLC_CELC,
  title={Improving multiple sclerosis lesion segmentation across clinical sites: A federated learning approach with noise-resilient training},
  author={Bai, L. and Wang, D. and Wang, H. and Barnett, M. and Cabezas, M. and Cai, W. and Calamante, F. and Kyle, K. and Liu, D. and Ly, L. and others},
  journal={Artificial Intelligence in Medicine},
  volume={152},
  pages={102872},
  year={2024},
  publisher={Elsevier}
}

@inproceedings{fedA3I,
  title={FedA3I: annotation quality-aware aggregation for federated medical image segmentation against heterogeneous annotation noise},
  author={Wu, N. and Sun, Zh. and Yan, Z. and Yu, L.},
  booktitle={AAAI Conference on Artificial Intelligence},
  volume={38},
  number={14},
  pages={15943--15951},
  year={2024}
}

@inproceedings{splitfed,
  title={Splitfed: When federated learning meets split learning},
  author={Thapa, Ch. and Arachchige, P. Ch. M. and Camtepe, S. and Sun, L.},
  booktitle={AAAI Conference on Artificial Intelligence},
  volume={36},
  number={8},
  pages={8485--8493},
  year={2022}
}

@inproceedings{Quality_adaptiveSplitFed,
  title={Quality-adaptive split-federated learning for segmenting medical images with inaccurate annotations},
  author={Kafshgari, Z. H. and Shiranthika, Ch. and Saeedi, P. and Baji{\'c}, I.},
  booktitle={20th International Symposium on Biomedical Imaging},
  pages={1--5},
  year={2023},
  organization={IEEE}
}

@inproceedings{kendall2018multi,
  title={Multi-task learning using uncertainty to weigh losses for scene geometry and semantics},
  author={Kendall, A. and Gal, Y. and Cipolla, R.},
  booktitle={Proceedings of the IEEE conference on computer vision and pattern recognition},
  pages={7482--7491},
  year={2018}
}

@article{consistency,
  title={Semi-supervised tissue segmentation from histopathological images with consistency regularization and uncertainty estimation},
  author={Sudhamsh, G. and Girisha, S. and Rashmi, R.},
  journal={Scientific Reports},
  volume={15},
  number={1},
  pages={6506},
  year={2025},
  publisher={Nature Publishing Group UK London}
}

@article{survey_noisy_label,
  title={Denoising and segmentation in medical image analysis: A comprehensive review on machine learning and deep learning approaches},
  author={Kumar, R. R. and Priyadarshi, R.},
  journal={Multimedia Tools and Applications},
  volume={84},
  number={12},
  pages={10817--10875},
  year={2025},
  publisher={Springer}
}

@inproceedings{FedLN,
  title={Federated Learning with Noisy Labels: Achieving Generalization in the Face of Label Noise},
  author={Tsouvalas, V. and Saeed, A. and {\"O}z{\c{c}}elebi, T. and Meratnia, N.},
  booktitle={First Workshop on Interpolation Regularizers and Beyond at NeurIPS 2022},
  year={2022}
}

@article{FedNoil,
  title={FedNOiL: a simple two-level sampling method for federated learning with noisy labels},
  author={Wang, Zh. and Zhou, T. and Long, G. and Han, B. and Jiang, J.},
  journal={arXiv preprint arXiv:2205.10110},
  year={2022}
}

@inproceedings{fedavg,
  title={Communication-efficient learning of deep networks from decentralized data},
   author={McMahan, B. and Moore, E. and Ramage, D. and Hampson, S. and y Arcas, B. A.},
  booktitle={Artificial intelligence and statistics},
  pages={1273--1282},
  year={2017},
  organization={PMLR}
}

@article{guan2024federated,
  title={Federated learning for medical image analysis: A survey},
  author={Guan, H. and Yap, P. T. and Bozoki, A. and Liu, M.},
  journal={Pattern Recognition},
  pages={110424},
  year={2024},
  publisher={Elsevier}
}

@article{Split_Learning,
  title={Split learning for health: Distributed deep learning without sharing raw patient data},
  author={Vepakomma, P. and Gupta, O. and Swedish, T. and Raskar, R.},
  journal={arXiv preprint arXiv:1812.00564},
  year={2018}
}

@article{fedDM,
  title={FedDM: Federated weakly supervised segmentation via annotation calibration and gradient de-conflicting},
  author={Zhu, M. and Chen, Zh. and Yuan, Y.},
  journal={IEEE Transactions on Medical Imaging},
  volume={42},
  number={6},
  pages={1632--1643},
  year={2023},
  publisher={IEEE}
}

@article{fedmix,
  title={FedMix: Mixed Supervised Federated Learning for Medical Image Segmentation},
  author={Wicaksana, J. and Yan, Z. and Zhang, D. and Huang, X. and Wu, H. and Yang, X. and Cheng, K. T.},
  journal={IEEE Transactions on Medical Imaging},
  year={2022},
  publisher={IEEE}
}

@article{FedNCL,
  title={Federated noisy client learning},
  author={Tam, K. and Li, L. and Han, B. and Xu, Ch. and Fu, H.},
  journal={arXiv preprint arXiv:2106.13239},
  year={2021}
}

@article{ARFL,
  title={Auto-weighted robust federated learning with corrupted data sources},
  author={Li, Sh. and Ngai, E. and Ye, F. and Voigt, Th.},
  journal={ACM Transactions on Intelligent Systems and Technology (TIST)},
  volume={13},
  number={5},
  pages={1--20},
  year={2022},
  publisher={ACM New York, NY}
}

@article{saeedi2017automatic,
  title={Automatic identification of human blastocyst components via texture},
  author={Saeedi, P. and Yee, D. and Au, J. and Havelock, J.},
  journal={IEEE Transactions on Biomedical Engineering},
  volume={64},
  number={12},
  pages={2968--2978},
  year={2017},
  publisher={IEEE}
}

@inproceedings{dice,
  title={Generalised dice overlap as a deep learning loss function for highly unbalanced segmentations},
  author={Sudre, Carole H and Li, Wenqi and Vercauteren, Tom and Ourselin, Sebastien and Jorge Cardoso, M},
  booktitle={Deep Learning in Medical Image Analysis and Multimodal Learning for Clinical Decision Support: Third International Workshop, DLMIA 2017, and 7th International Workshop, ML-CDS 2017, Held in Conjunction with MICCAI 2017, Qu{\'e}bec City, QC, Canada, September 14, Proceedings 3},
  pages={240--248},
  year={2017},
  organization={Springer}
}

@inproceedings{deeplabV3+,
  title={Encoder-decoder with atrous separable convolution for semantic image segmentation},
  author={Chen, L. Ch. and Zhu, Y. and Papandreou, G. and Schroff, F. and Adam, H.},
  booktitle={the European conference on computer vision (ECCV)},
  pages={801--818},
  year={2018}
}

@article{co_training,
  title={Deep co-training for semi-supervised image segmentation},
  author={Peng, J. and Estrada, G. and Pedersoli, M. and Desrosiers, Ch.},
  journal={Pattern Recognition},
  volume={107},
  pages={107269},
  year={2020},
  publisher={Elsevier}
}

@inproceedings{ResNet50,
  title={Deep residual learning for image recognition},
  author={He, K. and Zhang, X. and Ren, Sh. and Sun, J.},
  booktitle={Proceedings of the IEEE conference on computer vision and pattern recognition},
  pages={770--778},
  year={2016}
}

@article{PratondoChuiOng2016,
  title   = {Robust Edge-Stop Functions for Edge-Based Active Contour Models in Medical Image Segmentation},
  author  = {Pratondo, A. and Chui, Ch. K. and Ong, S. H.},
  journal = {IEEE Signal Processing Letters},
  year    = {2016},
  volume  = {23},
  number  = {2},
  pages   = {222--226},
  doi     = {10.1109/LSP.2015.2508039}
}

@inproceedings{Crete2007,
  title     = {The Blur Effect: Perception and Estimation with a New No-Reference Perceptual Blur Metric},
  author    = {Cr{\'e}t{\'e}, F. and Dolmiere, T. and Ladret, P. and Nicolas, M.},
  booktitle = {Proc. SPIE},
  year      = {2007}
}

@article{ma2020learning,
  title={Learning geodesic active contours for embedding object global information in segmentation CNNs},
  author={Ma, J. and He, J. and Yang, X.},
  journal={IEEE Transactions on Medical Imaging},
  volume={40},
  number={1},
  pages={93--104},
  year={2020},
  publisher={IEEE}
}

@inproceedings{abhishek2025can,
  title={What can we learn from inter-annotator variability in skin lesion segmentation?},
  author={Abhishek, K. and Kawahara, J. and Hamarneh, Gh.},
  booktitle={MICCAI Workshop on Deep Generative Models},
  pages={23--33},
  year={2025},
  organization={Springer}
}

@inproceedings{codella2018isic,
  title={Skin Lesion Analysis Toward Melanoma Detection: A Challenge at ISBI 2017},
  author={Codella, N. C. and Gutman, D. and Celebi, M. E. and Helba, B. and Marchetti, M. A. and Dusza, S. W. and Kalloo, A. and Liopyris, K. and Mishra, N. and Kittler, H. and Halpern, A.},
  booktitle={2018 IEEE 15th International Symposium on Biomedical Imaging (ISBI 2018)},
  pages={168--172},
  year={2018},
  organization={IEEE}
}

@inproceedings{abhishek2024segmentation,
  title={Segmentation Style Discovery: Application to Skin Lesion Images},
  author={Abhishek, K. and Kawahara, J. and Hamarneh, Gh.},
  booktitle={International Conference on Medical Image Computing and Computer-Assisted Intervention},
  pages={24--34},
  year={2024},
  organization={Springer}
}

\end{document}